\documentclass[aip,jcp,amsmath,amssymb,floatfix,reprint,citeautoscript]{revtex4-2}

\usepackage[utf8]{inputenc}
\usepackage{graphicx}
\usepackage{wrapfig}
\usepackage{natmove}
\usepackage{bibunits}
\usepackage[colorlinks,allcolors=black,citecolor=blue,urlcolor=blue]{hyperref}
\usepackage{chemformula}
\usepackage{physics}
\usepackage[debrief]{silence}
\graphicspath{{figures/}}

\defaultbibliography{paper}
\defaultbibliographystyle{aipnum4-2}

\begin{document}

\def\mytitle{Elucidating the Nature of $\pi$-hydrogen Bonding in Liquid Water and Ammonia}
\title{\mytitle}

\author{Krystof Brezina}

\author{Hubert Beck}

\author{Ondrej Marsalek*}
\affiliation{
Charles University, Faculty of Mathematics and Physics, Ke Karlovu 3, 121 16 Prague 2, Czech Republic
}

\date{\today}

\begin{abstract}

Aromatic compounds form an unusual kind of hydrogen bond with water and ammonia molecules, known as the $\pi$-hydrogen bond.
In this work, we report \textit{ab initio} path integral molecular dynamics simulations enhanced by machine-learning potentials to study the structural, dynamical, and spectroscopic properties of solutions of benzene in liquid water and ammonia.
Specifically, we model the spatial distribution functions of the solvents around the benzene molecule, establish the $\pi$-hydrogen bonding interaction as a prominent structural motive, and set up existence criteria to distinguish the $\pi$-hydrogen bonded configurations.
These serve as a structural basis to calculate binding affinities of the solvent molecules in $\pi$hydrogen bonds, identify an anticooperativity effect across the aromatic ring in water (but not ammonia), and estimate $\pi$-hydrogen bond lifetimes in both solvents.
Finally, we model hydration-shell-resolved vibrational spectra to clearly identify the vibrational signature of this structural motif in our simulations.
These decomposed spectra corroborate previous experimental findings for benzene in water, offer additional insights, and further emphasize the contrast between $\pi$-hydrogen bonds in water and in ammonia.
Our simulations provide a comprehensive picture of the studied phenomenon and, at the same time, serve as a meaningful \textit{ab initio} reference for an accurate description of $\pi$-hydrogen bonding using empirical force fields in more complex situations, such as the hydration of biological interfaces.

\end{abstract}

{\maketitle}

\begin{bibunit}

\section{Introduction}
\label{sec:intro}

A $\pi$-hydrogen bond is a hydrogen bonding interaction formed between a partially positively charged hydrogen atom of a donor molecule and the $\pi$-electron density of an aromatic acceptor.
Such interactions have been observed with a variety of donor molecules including halogen hydrides~\cite{Nekoei2019/10.1039/C8CP07003B}, aliphatic hydrogens,~\cite{Nishio2011/10.1039/C1CP20404A} alcohols,~\cite{Mons2000/10.1021/jp993178k} liquid ammonia~\cite{Brezina2020/10.1021/ACS.JPCLETT.0C01505} and water~\cite{Suzuki1992/10.1126/SCIENCE.257.5072.942, Gierszal2011/10.1021/JZ201373E}.
The interaction of aromatics with liquid water (\ch{H2O}) bears everyday-life pertinence due to its role in the solvation of aromatic residues at biological interfaces, structural stabilization inside proteins~\cite{Steiner2001/10.1006/JMBI.2000.4301}, as well as contamination of sea waters by oil-derived aromatic substances dissolved during oil spills.~\cite{Gros2014/10.1021/ES502437E}
For instance, the prototypical aromatic compound benzene has a surprisingly high solubility of up to 20~mM in liquid water at ambient conditions.~\cite{Arnold1958/10.1021/I460004A016}
Chemically similar to water, liquid ammonia (\ch{NH3}) is a noticeably better solvent for aromatics~\cite{Schewe2022/10.1021/ACS.JPCB.1C08172} and such mixtures are significant from the synthetic and industrial viewpoint as an entryway to Birch chemistry.~\cite{Birch1996/10.1351/PAC199668030553}

In general, water and ammonia in their liquid state are considered to be similar based on their molecules both being simple first-period, isoelectronic hydrides, as well as on the formation of hydrogen-bonded structures.
However, because of the less electronegative nature of the nitrogen atom in ammonia, the hydrogen bond strength is substantially lowered when compared to water, which causes numerous remarkable differences between the two liquids.~\cite{Boese2003/10.1063/1.1599338, Krishnamoorty2022/10.1021/ACS.JPCLETT.2C01608}
At the macroscopic level, this can represented, for instance, by the large difference in their boiling points of 133~K despite their nearly identical molar masses.~\cite{Glazier2010/10.1021/ED100691N}
At the molecular level, the different hydrogen bond strength manifests in various properties of the two liquids, including both structural and dynamic quantities such as for example, infrared (IR) vibrational spectra.
Here, water exhibits a pronounced red shift of the OH stretch in comparison to the gas phase from approximately 3700 to 3400 cm$^{-1}$ owing to the softening of the effective bond potential in the presence of the hydrogen bond acceptor.~\cite{Kosower2011/10.1039/C1RA00443C}
On the other hand, ammonia remains nearly unchanged in this aspect.~\cite{Slipchenko2008/10.1063/1.2884927/977357}

To study the nature of $\pi$-hydrogen bonds in these protic solvents, limiting oneself to benzene as the prototypical aromatic species gives a general picture of the phenomenon.
In this regard, a key experimental contribution reports the measurement of the multivariate-curve resolution Raman spectrum~\cite{Perera2008/10.1021/JA077333H} of benzene in water.~\cite{Gierszal2011/10.1021/JZ201373E} 
In this study, the authors present the so-called solute-correlated (SC) spectrum, which represents a contribution to the overall Raman intensity that captures the vibrational character of the benzene solute as well as that of its immediate surroundings.
This reveals an unexpected vibrational feature at approximately 3610~cm$^{-1}$, which is interpreted as an imprint of the effect of the water molecules engaging in $\pi$-hydrogen bonds on the solute vibrations.~\cite{Gierszal2011/10.1021/JZ201373E}
The particular frequency of the observed $\pi$-hydrogen bond peak suggests that water molecules participating in a $\pi$-hydrogen bond are vibrationally blue-shifted compared to water molecules in the liquid bulk and, as such, have more of a character of an isolated molecule since they are not fully involved in the water--water hydrogen bond network.
Our recent work focusing on the corresponding liquid ammonia solutions using \textit{ab initio} molecular dynamics (AIMD) with dispersion-corrected hybrid density functionals finds the equivalent $\pi$-hydrogen bond spectral feature in the IR SC spectrum of benzene obtained by exact decomposition of the dipole--dipole autocorrelation function Fourier transforms.~\cite{Brezina2020/10.1021/ACS.JPCLETT.0C01505}
Additionally, the blue shift of water has been given a degree of computational support at the level of individual OH bond frequency analysis in AIMD simulations using the BLYP density functional~\cite{Becke1988/10.1103/PhysRevA.38.3098, Lee1988/10.1103/PhysRevB.37.785} to describe the condensed-phase system.~\cite{Mallik2008/10.1021/JP801405A, Choudhary2015/10.1021/ACS.JPCB.5B03371}

The solvation structure of benzene at the atomistic level has been studied using MD simulations~\cite{Allesch2007/10.1021/jp065429c, Allesch2008/10.1063/1.2806288/928000} at the PBE level of theory~\cite{Perdew1996/10.1103/PhysRevLett.77.3865} in water and, as part of our previous work~\cite{Brezina2020/10.1021/ACS.JPCLETT.0C01505}, using the revPBE0-D3 density functional~\cite{Perdew1996/10.1103/PhysRevLett.77.3865, Zhang1998/10.1103/PhysRevLett.80.890, Adamo1999/10.1063/1.478522, Goerigk2011/10.1039/c0cp02984j} in liquid ammonia.
Both works come to a qualitatively comparable conclusion concerning the shape of the spatial distribution function (SDF) which suggests a substantial degree of similarity of structural aspects in both solvents.
The reported SDFs have a geoid-like shape with pronounced maxima in the pole regions above and below the planar benzene molecule, where $\pi$-hydrogen bonds were found to occur, implying a major role of $\pi$-hydrogen bonds as a binding motive in benzene solvation.

All of the above-mentioned simulations employ a classical approximation for the atomic nuclei.
However, a $\pi$-hydrogen bond, just like any other hydrogen bond, is an interaction that features the very light hydrogen nucleus and, therefore, can be expected to exhibit pronounced nuclear quantum effects (NQEs) even at ambient conditions.~\cite{Li2011/10.1073/PNAS.1016653108, Markland2018/10.1038/s41570-017-0109}
The overall change in the character of a particular hydrogen bond upon inclusion of quantum nuclei into the picture is derived from a fine balance between the so-called competing NQEs, which can have a two-way effect of strengthening already strong hydrogen bonds, but also weakening initially weak ones.~\cite{Habershon2010/10.1063/1.3167790/938289, Li2011/10.1073/PNAS.1016653108}
This can have profound manifestations in the structure and dynamics of a hydrogen-bonded system at the local level but can also act at longer scales by affecting, for instance, the cooperativity of hydrogen bonds.
As such, the multiple research works up to this point provide a qualitative picture of various isolated aspects of $\pi$-hydrogen bonds, but a rounded computational treatment with highly accurate electronic structure and the inclusion of quantum nuclei is missing.
At the same time, having access to an accurate \textit{ab initio} reference is important as $\pi$-hydrogen bonding is prominent in biological systems, but common force fields were found to be unreliable for this purpose~\cite{Allesch2008/10.1063/1.2806288/928000}.

To provide a comprehensive picture of the phenomenon, we perform thermostatted ring polymer molecular dynamics~\cite{Rossi2014/10.1063/1.4883861/73314} (TRPMD) simulations of dilute solutions of benzene in liquid water (at 300~K) and ammonia (at 223~K) using revPBE0-D3 hybrid density functional theory to model the interaction potential.
Such simulation exploits the classical isomorphism of the imaginary-time path-integral formalism, where quantum particles are represented as harmonically-coupled ring polymers of replicas of the classical system.
This allows for the use of classical AIMD techniques at a computational cost elevated linearly with the number of path-integral replicas to address NQEs of both structural and dynamical properties of studied systems.
Performing these simulations naively using the traditional on-the-fly electronic structure calculation approach to AIMD is possible, but obtaining long enough trajectories for tightly converged statistical properties is impractical due to the increased cost of the path integral simulations.
Therefore, we approach this issue mainly by combining the traditional approach with Behler--Parrinello high-dimensional neural network potentials~\cite{Behler2007/10.1103/PHYSREVLETT.98.146401, Schran2020/10.1063/5.0016004} (NNPs) to fit the reference \textit{ab initio} potential energy surface and perform 2~ns long simulations at a drastically reduced computational cost while maintaining the accuracy of the simulations. 
The shorter 250~ps \textit{ab initio} trajectories themselves are used for the training of the NNPs, their validation, and the calculation of molecular dipole moments, which are not directly accessible through NNPs.
A comprehensive summary of the employed computational methodology is available in Section~\ref{si-sec:comp-details} of the Supporting Information.
With this methodology, we address the molecular-level solvation structure, discuss the $\pi$-hydrogen bond orientation, cooperativity, lifetimes, and, finally, vibrational spectroscopy, where we provide theoretical insight with immediate relevance for the interpretation of the underlying experimental work.

\section{Results and Discussion}

In the following paragraphs, we present and discuss the results obtained in this work.
In Section~\ref{sec:structure}, we analyze the molecular solvation structure with focus on the geometry of the $\pi$-hydrogen bonds.
This allows us to set up a time-dependent existence criterion to assess whether a solvent molecule is or is not participating in a $\pi$-hydrogen bond at a given instant, which is then used in Section~\ref{sec:cooperativity} to address the issue of $\pi$-hydrogen bond cooperativity across the two binding sites that the benzene molecule offers.
In Section~\ref{sec:dynamics}, we exploit the existence criterion again to look into the kinetics of $\pi$-hydrogen bond formation and, finally, we look into the problem of vibrational spectroscopy in Section~\ref{sec:vibrations} by modeling the influence of the benzene solute on the vibrational spectra from \textit{ab initio} data, finding an effect consistent with the above experimental observations and connecting its origin to vibrational features in the solvent part of the spectrum mainly due to $\pi$-hydrogen bonded solvent molecules.

\subsection{Solvation structure}
\label{sec:structure}

\begin{figure}[tb!]
    \centering
    \includegraphics{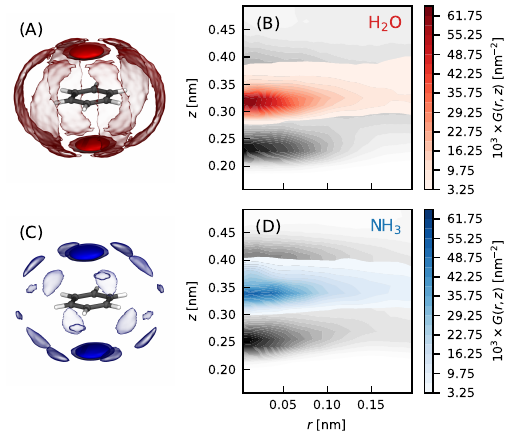}
    \caption{
    Spatial distribution functions of water (top row) and ammonia (bottom row) around the benzene solute.
    Panel A shows two different contours of the 3D SDF $G(r, z, \varphi)$ for the oxygen atoms of water at 95~$\mathrm{\AA}^{-3}$ (transparent) and 130~$\mathrm{\AA}^{-3}$ (solid).
    Panel B shows the same SDF integrated around the polar angle $\varphi$, zoomed in on the $\pi$-hydrogen bond cap regions and resolved for the oxygen atoms (red) and the hydrogens (gray).
    Panels C and D show the same data for the ammonia simulations with the nitrogen-related quantities colored blue.
    }
    \label{fig:sdfs}
\end{figure}

\begin{figure*}[tb!]
    \centering
    \includegraphics{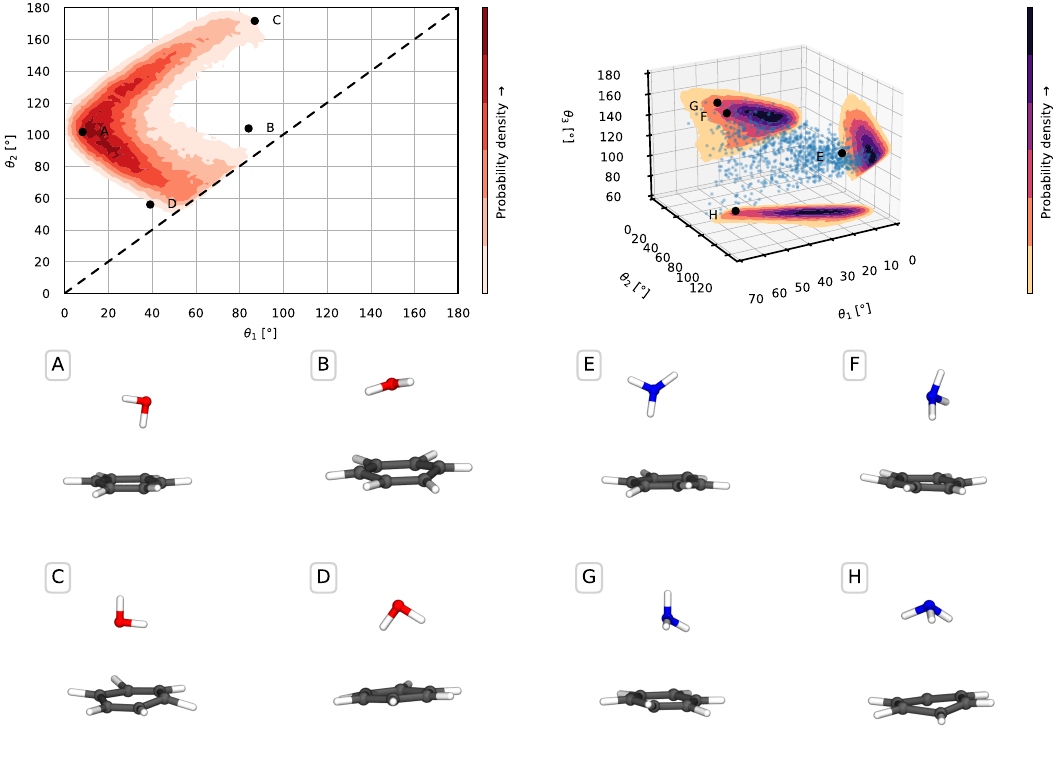}
    \caption{
    Analysis of the orientation of solvent molecules around the benzene solute.
    Top left panel: the orientation of water molecules is described by two angles, $\theta_1$ and $\theta_2$ which describe the angle between the O--H bond and the solute normal with $\theta_1$ describing the hydrogen that is closer to the solute and $\theta_2$ the further one.
    Both angles reach 0$^\circ$ when the bond is oriented towards the solute with the hydrogen atom.
    Panels A, B, C, and D: snapshots of representative orientations of water molecules that occur in the cylinder defined by the criterion $h_i^{(k)}(t)$.
    Only panel A is clearly an oriented $\pi$-hydrogen bond.
    Top right panel: identical type of plot for ammonia (as for water on the left) with the addition of the angle $\theta_3$ describing the orientation of the third hydrogen atom.
    The 3D distribution itself is shown as a scatter plot of a subset of points selected from the simulations.
    The 2D marginal distributions capturing the correlations of the orientation of two different N--H bonds are shown on the corresponding coordinate planes.
    Panels E, F, G, and H: representative configurations of solvent molecules in the $\pi$-hydrogen bond cylinder.
    }
    \label{fig:orientation}
\end{figure*}

A global picture of the solvation of benzene by water and ammonia is encoded in the SDF $G(r, z, \varphi)$ of the solvent around the solute.
The SDF is a 3D histogram of the positions of a selected atomic species described conveniently in this case in standard cylindrical coordinates $r$, $z$, and $\varphi$ with the origin located in the solute center of mass and the plane of the solute aligned with $z = 0$.
The SDFs for both solvents were obtained as statistical averages of heavy-atom positions over the 1~ns long C-NNP trajectories and the resulting data is shown in Figure~\ref{fig:sdfs}, in panel A for water and in panel C for ammonia.
Both SDFs share similar features and thus will be discussed concurrently.
Their overall shape can be compared to that of a rotational ellipsoid with the shorter axis in the $z$ direction and they comprise two distinct regions: a set of hexagonally arranged stripe-like lobes around the solute plane that feature a hydrophobic solvation regime~\cite{Ben-Amotz2016/10.1146/ANNUREV-PHYSCHEM-040215-112412} (this issue will not be discussed in detail in this work) and two caps at the poles surrounding $r = 0$ which represent global maxima of the SDF and which will be identified as the regions in which $\pi$-hydrogen bonds are formed.
Benzene thus simultaneously offers two binding sites for solvent molecules to form $\pi$-hydrogen bonds that we will denote for later use as site $A$ (above the ring, $z > 0$) and site $B$ (under the ring, $z < 0$).
The only noticeable difference between the two SDFs is the less structured hydrophobic part of the ammonia SDF than that of the water SDF; the $\pi$-hydrogen bond caps look comparable.
This suggests that the solvation structure of the benzene solute is similar in the two solvents, and the slightly stronger polarity of water molecules does not play a structure-defining role in this instance.
Note that the data shown in panel C of Figure~\ref{fig:sdfs} is directly comparable to our previous data shown in Reference~\citenum{Brezina2020/10.1021/ACS.JPCLETT.0C01505}.
However, unlike in our previous work, where the \textit{ab initio} data required a smoothing procedure, the C-NNP data here is presented raw and originates from TRPMD simulations.

More insight into the nature of the $\pi$-hydrogen bond caps is achieved by calculating a similar SDF for the hydrogen atoms. For graphical clarity, we integrate the 3D SDFs around the polar angle $\varphi$ and show the resulting 2D SDFs $G(r, z)$ for both the heavy atoms and the hydrogens in panels B and D of Figure~\ref{fig:sdfs}, respectively, for the two solvents and averaged over both binding sites $A$ and $B$.
Note that the shown extent focuses on the $\pi$-hydrogen bond caps only.
Here, each colored, heavy-atom peak has a gray, hydrogen peak approximately 0.8~\AA\ under it, which suggests that an oriented interaction \ch{C6H6} $\cdots$ HX (X = O, N) is formed where the XH bond faces the solute along the $z$ axis with its hydrogen pointing towards it.
This interaction, prominently featured in the simulation data in both solvents, is the $\pi$-hydrogen bond.

Next, we want to explore how are the two independent 2D SDF clouds correlated, or, in other words, what possible orientations of $\pi$-hydrogen bonds occur in our simulations.
For that purpose, we define an existence criterion for $\pi$-hydrogen bonds which picks all molecules that are located inside a cylinder that isolates the $\pi$-hydrogen bond caps from the rest of the SDFs.
This criterion is a time series over the simulated trajectory and is defined individually for each $n$-th solvent molecule and $k$-th imaginary time replica as
\begin{equation}
\label{eq:existence-criterion-mol}
\begin{split}
    h_{nk}^{(A)}(t)
    &=
    \Theta\left[r_{nk}(t) - r_0\right] \Theta\left[z_{nk}(t) - z_0\right] \\
    &=
    \begin{cases}
        1 \quad \text{in a $\pi$-hydrogen bond} \\ 0 \quad \text{out of $\pi$-hydrogen bond}
    \end{cases}
\end{split}
\end{equation}
for binding site $A$ and identically as
\begin{equation}
    h_{nk}^{(B)}(t)
    =\Theta\left[r_{nk}(t) - r_0\right] \Theta\left[z_{nk}(t) + z_0\right]
\end{equation}
for binding site $B$.
In both definitions, $\Theta$ is the step function, and $r_{nk}$ and $z_{nk}$ are the cylindrical coordinates of the heavy atom X of a given molecule and replica.
The parameters $r_0$ and $z_0$ are selected based on the shape and extent of the individual SDFs: for water we set $r_0 = 0.12$~nm and $z_0 = 0.40$~nm and for ammonia $r_0 = 0.14$~nm and $z_0 = 0.41$~nm.
Next, we define a set of angles $\theta_j$ for each solvent molecule ($j=1,2$ for water and $j=1,2,3$ for ammonia) which denote the angles between each HX bond and the $z$ axis where a 0$^\circ$ angle is defined as the HX bond being collinear with $z$ and with the hydrogen atom pointing towards the solute. All angles are then ordered in $j$ by the vertical distance of the hydrogen atom from the solute so that $\theta_1$ is the angle of the $\pi$-hydrogen bond.
The distributions of these angles for replicas that comply with $h_{nk}^{(S)}(t)$ are shown in Figure~\ref{fig:orientation} averaged over both binding sites $S = A, B$.
For water (Figure~\ref{fig:orientation}, top panel), we observe a V-shaped distribution with a maximum in $\theta_1 \sim 10^\circ$ and $\theta_2 \sim 110^\circ$.
This corresponds to the expected, albeit slightly tilted orientation of the $\pi$-hydrogen bond with the solvent molecule pointing one of its hydrogen towards the solute: the value of $\theta_2$ here derives from the equilibrium bond angle in water of $104.5^\circ$ (Figure~\ref{fig:orientation}, snapshot A).
The tilt originates as a result of the contribution of the orientational entropy to the shown population: since there is only a single way to realize the perfectly oriented $\theta_1 = 0^\circ$ configuration, this is practically not seen during the simulation.
The fact that it is indeed the energetically most favored configuration is shown by factoring out the entropic sine contributions as shown in Section~\ref{si-sec:additional_results} of the Supporting information.
However, the extent of the distribution points to the fact that a richer pool of configurations is encountered in the simulations than just the idealistic $\pi$-hydrogen bond.
In fact, the selection of molecules that comply with $h_{nk}^{(S)}(t)$ also includes molecules that are not in a $\pi$-hydrogen bond at all and just happen to be located in the region of space, but likely engaging in regular water--water hydrogen bond network (Figure~\ref{fig:orientation}, snapshots B, C and, as a borderline case, D).
Therefore, based on this analysis, we update the selection criterion to also reflect the $\pi$-hydrogen bond angle:
\begin{equation}
    \label{eq:existence-criterion}
    h_{nk}^{(S)}(t)
    \mathrel{{^*}{=}}
    \Theta \left[ \theta_{1,nk}(t) - \theta_{1,0} \right],
\end{equation}
where the cutoff angle was set to $\theta_1 = 40^\circ$ for both solvents.
A discussion along the same lines arises for ammonia. 
The 3D distribution in the corresponding angles is shown in the top right panel of Figure~\ref{fig:orientation} and representative configurations in panels E, F, G, and H of the same Figure. 
Here, we remark that the preferred $\pi$-hydrogen bond configuration is also oriented vertically towards the solute with a slightly larger tilt of $\sim 20^\circ$.

\begin{figure}[tb!]
    \centering
    \includegraphics{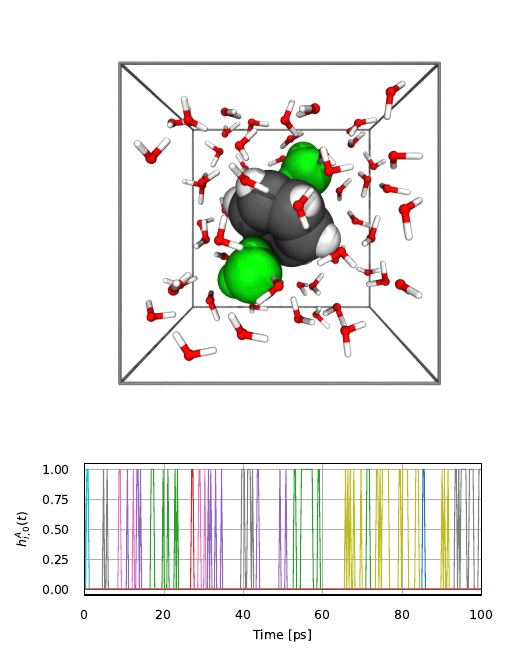}
    \caption{
    An illustration of the $\pi$-hydrogen bond existence time series.
    Snapshot: a single simulation frame showing the simulated periodic system consisting of water solvent and benzene solute for a single path-integral replica.
    Water molecules that fulfill the existence criterion are highlighted in the van der Waals representation and in green.
    Bottom panel: A graphical representation of the time dependence of $h_{n, k=0}^{A}(t)$ (Equation~\ref{eq:existence-criterion-mol}) over a 100~ps long segment of one the simulated C-NNP path-integral trajectories and for a single replica.
    The molecular index $n$ is represented by the different colors of the curves, showcasing the sampling of the exchange of molecules in the binding site $A$.
    }
    \label{fig:criterion}
\end{figure}

\subsection{$\pi$-hydrogen bond cooperativity}
\label{sec:cooperativity}

As established above, the benzene molecule can engage in two $\pi$-hydrogen bonds at a time by binding a solvent molecule in sites $A$ and $B$.
Here, we ask what the populations of the zero-, one- and two-bonded configurations are in the simulated trajectories and to what extent the two coexisting hydrogen bonds cooperate or whether they can be considered independent.

\begin{figure}[tb!]
    \centering
    \includegraphics{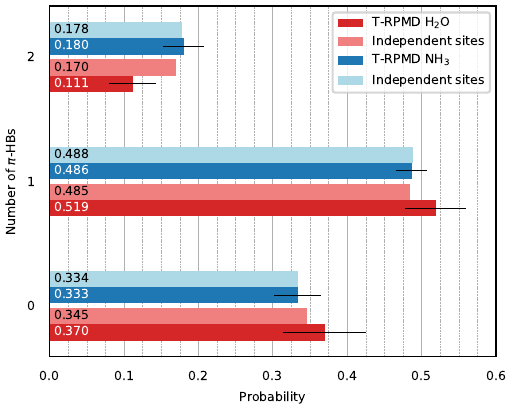}
    \caption{
    Probabilities of finding the zero-, one- and two-bonded configurations within our simulated trajectories.
    The darker bars show real, observed populations.
    The lighter bars then show probabilities that are calculated based on the premise of binding site independence.
    The error bars have been estimated using the block averaging method individually for each $\delta[h_k(t) - N]$ time series.
    }
    \label{fig:populations}
\end{figure}

To begin such analysis, we calculated the $\pi$-hydrogen bond existence criterion for the solute in our \textit{ab initio} simulations (not C-NNP, see below) by summing over the time series of individual solvent molecules and sites as
\begin{equation}
    h_k(t)
    =
    \sum_{n, S} {}h_{nk}^{(S)}(t)
    =
    \begin{cases}
        0 \text{ $\pi$-hydrogen bonds} \\
        1 \text{ $\pi$-hydrogen bond} \\
        2 \text{ $\pi$-hydrogen bonds.}
    \end{cases}
\end{equation}
The probabilities for the individual species ($p(0)$ for 0 $\pi$-hydrogen bonds, $p(1)$ for a single $\pi$-hydrogen bond, and $p(2)$ for both binding sites occupied) were calculated as a simple average over the time variable and the $P$ path integral replicas
\begin{equation}
    p(N)
    =
    \frac{1}{PT} \int_0^T \dd t \ \sum_{k=1}^{P} \delta\left[h_k(t) - N\right]
\end{equation}
for each $N = 0, 1, 2$.
The obtained numbers $p(N)$ are shown in Figure~\ref{fig:populations} using the darker bars.
In both solvents, the most likely configuration is that with only a single $\pi$-hydrogen bond with 51.9~\% in water and 48.6~\% in ammonia.
This is followed by the configuration with no $\pi$-hydrogen bonds populated at 37.0~\% in water and 33.3~\% in ammonia, and the least probable configuration is the one with the maximal occupancy of 2 $\pi$-hydrogen bonds with 11.1~\% in water and 18.0~\% in ammonia.
The probabilities $p(N)$ are related to the probabilities of occupying the specific binding sites $A$ and $B$ as follows (for this purpose, we borrow the notation $A$ and $B$ for referring to the sets of microstates corresponding to a solvent molecule being bound at the given binding site).
By the symmetry of the problem, we clearly have $p(A) = p(B)$.
Then, $p(2) = p(A \cap B) = p(A)p(A|B)$, $p(1) = p(A \cup B) - p(A \cap B) = 2p(A) - p(2)$, and, by completeness, $p(0) = 1 - p(1) - p(2)$.

The possible cooperativity was examined by comparing the observed populations to a model that assumes the independence of both binding sites.
This model was constructed based on the premise that independent occurrences of binding to sites $A$ and $B$ (the corresponding probabilities are denoted as $p_\mathrm{ind}$) comply with $p_\mathrm{ind}(A|B) = p_\mathrm{ind}(A)$, which implies that $p_\mathrm{ind}(2) = p_\mathrm{ind}^2(A)$.
This leads to a single equation for two unknown independent probabilities, $p_\mathrm{ind}(0)$ and $p_\mathrm{ind}(1)$, giving us the freedom to fix one of these quantities.
In this work, we choose to preserve the free-energy difference between states 0 and 1, which requires setting $p_\mathrm{ind}(1) / p_\mathrm{ind}(0) = p(1) / p(0)$.
This allows us to estimate all the independent probabilities $p_\mathrm{ind}(N)$.
These results are shown in Figure~\ref{fig:populations} using the lighter shading next to the bars for the true observed probabilities $p(N)$.
For benzene in water, comparing the real and independent probabilities is curious because it points to an anticooperativity effect between the two binding sites: the observed 11.1~\% population of the double-bonded state is significantly lower compared to the value of 17.0~\% predicted by the independent model.
We explain this observation by considering the stronger impact a water molecule has on the solute, drawing the $\pi$-electron density of the aromatic ring to its side upon the formation of a $\pi$-hydrogen bond.
This way, it weakens the other binding site and makes it less prone to forming a second $\pi$-hydrogen bond at the same time.
In ammonia, this effect does not arise, and the independent model represents an excellent approximation of the real system.
An illustration of this effect on electron density in gas-phase clusters consisting of a single benzene molecule and two $\pi$-hydrogen bonded solvent molecules is presented in Section~\ref{si-sec:additional_results} of the Supporting information.

Finally, we comment on the performance of C-NNP models in the simulation of $\pi$-hydrogen bond cooperativity.
Our initial attempts to investigate the $\pi$-hydrogen bond cooperativity in C-NNP trajectories revealed that the anti-cooperativity effect for aqueous systems is reproduced poorly by the committee of Behler--Parrinello (BP) NNPs --- it is considerably lower than in the original \textit{ab initio} trajectories.
$\pi$-hydrogen bond cooperativity is a very subtle effect and, hence, more difficult to accurately reproduce than other properties such as spatial distribution functions or vibrational spectra.
A quantitative investigation into the accuracy of the employed models can be found in Section~\ref{si-sec:additional_results} the Supporting information and shows that the force errors are below the threshold typically needed to reproduce most standard properties.
One challenge for the C-NNP description of $\pi$-hydrogen bonds is that the typical distance between the center of the benzene ring and the oxygen atom of a water molecule is between 3.0 and 3.5~\AA\ (as shown in Figure~\ref{fig:sdfs}).
This suggests that in configurations with two simultaneous $\pi$-hydrogen bonds, the water molecule at the opposite site is at the very edge of the neighborhood cutoff radius of the BP NNP, which we set, as per usual, to 12~$a_0$ ($\approx$ 6.35~\AA).
Since the resolution of the atom-centered symmetry functions towards the edge of the cutoff radius is poor, an accurate prediction of this long-range effect is unlikely by these local MLPs.
The more advanced equivariant message-passing graph NNPs such as NequIP can mitigate this downside, as the message-passing architecture extends the effective field of view beyond the set cutoff radius.~\cite{Schutt2021/10.48550/arXiv.2102.03150}
For this reason, we trained a NequIP model~\cite{Batzner2022/10.1038/s41467-022-29939-5} for the classical benzene--water system and ran the same MD simulations with it as well.
However, we observed no improvement in these trajectories, as the anti-cooperativity effect remains severely underestimated.
Further research will be necessary to disentangle the reasons why both BP and equivariant NNPs fail to reproduce this effect quantitatively.
One plausible explanation lies in the training data set, which has been selected using query by committee (QbC) with BP NNPs and was used to train both model architectures.
It is possible that due to the short-range nature of BP NNPs, structures required for an accurate description of the anti-cooperativity effect have been left out of the training data set and that only a NequIP-specific QbC will lead to an NNP that reproduces all aspects of this system correctly.

\subsection{$\pi$-hydrogen bond lifetimes}
\label{sec:dynamics}

\begin{figure}[tb!]
    \centering
    \includegraphics{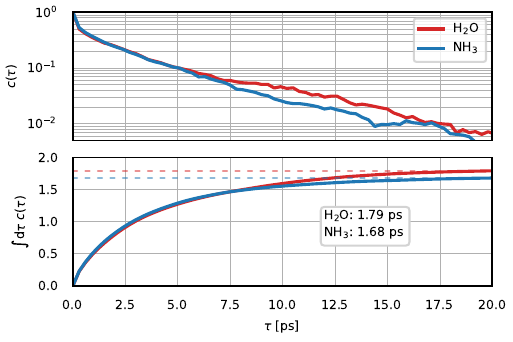}
    \caption{
    Time ACFs of the $\pi$-hydrogen bond existence criterion.
    The top panel shows the ACF for both water and ammonia on the interval of lag time from 0 to 20~ps on a logarithmic scale.
    The bottom panel shows the integrated ACFs to estimate the correlation times as the long time limit of these ACFs --- this is shown as the dashed lines.
    }
    \label{fig:acfs}
\end{figure}

The time dependence of the solvent existence criteria $h_{nk}^{(S)}(t)$ (Equation~\ref{eq:existence-criterion}) can be used to estimate the characteristic lifetime of the $\pi$-hydrogen bond.
To calculate the time correlation function, we first evaluate the RPMD time-dependent observable~\cite{Craig2005/10.1063/1.1777575} as
\begin{equation}
    h_{n}^{(S)}(t)
    =
    \frac{1}{P} \sum_{k=1}^{P} h_{nk}^{(S)}(t)
\end{equation}
by averaging over all the path-integral replicas for each molecule and each site separately.
This quantity no longer has discreet values between 0 and 1 but can acquire a continuum of values between these two bounds: this can be interpreted as the quantum-delocalized solvent molecule only partially fulfilling the $\pi$-hydrogen bond existence criterion due to its replicas entering and leaving the relevant region.
The autocorrelation function is then 
\begin{equation}
    c(\tau)
    =
    \frac{1}{2N_\mathrm{sol}} \sum_{n, S} \frac{\langle h_{n}^{(S)}(t_0) h_{n}^{(S)}(t_0 + \tau) \rangle}{\langle [ h_{n}^{(S)}(t_0) ]^2 \rangle}    
\end{equation}
and, as such, gives the probability that a $\pi$-hydrogen bond still exists after a delay $\tau$ if it initially existed at time $t_0$.
Such autocorrelation functions for both solvents are shown in Figure~\ref{fig:acfs} over the range of 20~ps and feature a typical shape with a sharper decrease at very short delays followed by a less abrupt, long, and monotonously decaying exponential regime.
Integration of these functions over the time variable reveals the time scale at which these correlations decay, which can be interpreted as the characteristic lifetime of $\pi$-hydrogen bonds.
Interestingly, these lifetimes are comparable in both solvents: for water, we obtained an autocorrelation time of approximately 1.8~ps, while for ammonia, a just slightly shorter value of 1.7~ps.
We propose that this agreement is coincidental as the solutions are at different temperatures and have different viscosities.
To set the ground for comparison, we additionally calculated the hydrogen bond lifetimes for the neat solvents as well using preexisting trajectories: the lifetime of water--water hydrogen bonds is approximately 3.9~ps and ammonia--ammonia hydrogen bonds approximately 1.4~ps.
This indirectly suggests that solvent--solvent hydrogen bonds are stronger than $\pi$-hydrogen bonds in water but comparable or even ever so slightly weaker in ammonia.

\subsection{Vibrational spectroscopy}
\label{sec:vibrations}

\begin{figure}[tb!]
    \centering
    \includegraphics{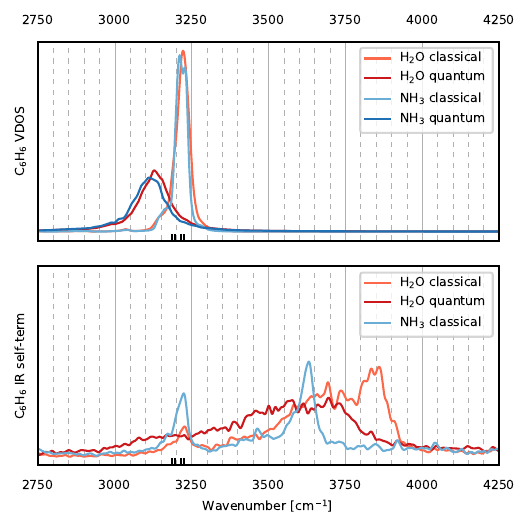}
    \caption{
    Vibrational characterization of the benzene solute in liquid water (light red for classical and dark red for TRPMD trajectories) and ammonia (light blue for classical and dark blue for TRPMD trajectories).
    Top panel: Vibrational density of states of the solute carbon and hydrogen atoms combined.
    Bottom panel: IR absorption self-term due to the benzene molecular dipole moment.
    Harmonic frequencies of gas-phase benzene calculated at the revPBE0-D3 hybrid DFT level are shown as black ticks at the bottom of each panel.
    }
    \label{fig:vibrations}
\end{figure}

Finally, we turn our attention to the simulation of vibrational spectra of both solutions in order to relate to the previous experimental findings~\cite{Gierszal2011/10.1021/JZ201373E} and to gain insight into the influence of the aromatic solute on the solvent.
In this regard, we present a qualitative perspective based on vibrational densities of states (VDOS) and infrared (IR) intensities.
VDOS is defined through the Fourier transform of the autocorrelation function of atomic velocities $\mathbf{v}_i$,
\begin{equation}
    \begin{split}
        C_{vv, i}(t)
        &=
        \langle \mathbf{v}_i(t_0) \cdot \mathbf{v}_i(t_0 + t) \rangle_{t_0}, \\
        I_\mathrm{VDOS}(\omega)
        &\propto
        \sum_{i} \int \dd t \ e^{-\mathrm{i} \omega t} C_{vv, i}(t),
    \end{split}
\end{equation}
where we write the total VDOS as a sum of atomic contributions.
While it has no experimentally measurable counterpart, it is the most direct and local probe into the mechanical vibrations of molecules.
Similarly, IR intensities, which correspond to experimentally measurable IR signals, can be related to the Fourier transform of the autocorrelation function of the total dipole moment $\mathbf{M}$ of the studied system,
\begin{equation}
\label{eq:ir}
    \begin{split}
        C_{MM}(t)
        &=
        \langle \mathbf{M}(t_0) \cdot \mathbf{M}(t_0 + t) \rangle_{t_0}, \\
        I_\mathrm{IR}(\omega)
        &\propto
        \omega^2 \int \dd t \ e^{-\mathrm{i} \omega t} C_{MM}(t).
    \end{split}
\end{equation}
Given a partitioning method to decompose the total dipole to molecular contributions as $\mathbf{M} = \sum_n \pmb{\mu}_n$, the total IR intensity can also be decomposed to contributions of pairs of molecules as
\begin{equation}
    I_\mathrm{IR}(\omega)
    =
    \omega^2 \sum_{n, m} \int \dd t \ e^{-\mathrm{i} \omega t} \langle \pmb{\mu}_n(t_0) \cdot \pmb{\mu}_m(t_0 + t) \rangle_{t_0},
\end{equation}
where not only the molecular self-terms ($n = m$), but also cross-correlation terms ($n \neq m$), contribute to the total intensity.
In the following analysis, the quantum vibrational spectra are considered to be the primary outcome, but we will frequently relate to the corresponding classical simulations, owing to the insight they provide into the interpretation of the presented spectral curves.

To motivate the discussion, we first inspect the vibrational character of the solute in both solvents.
The solute component of the VDOS in the region of benzene CH stretch modes between 2750 and 4250~cm$^{-1}$ is shown in the top panel of Figure~\ref{fig:vibrations} for water as well as ammonia.
Classically, the CH vibrations present in the VDOS as a relatively narrow peak located just above 3200~cm$^{-1}$ with an almost identical shape in both systems.
Its position closely corresponds to the position of gas-phase harmonic frequencies of the isolated benzene molecule (Figure~\ref{fig:vibrations}), suggesting that the vibrational motion of the solute is not strongly affected by its solvation.
The TRPMD VDOS exhibits a typical vibrational manifestation of NQEs, where the quantum spectrum is, compared to the classical one, broader and redshifted --- in this case by approximately 100~cm$^{-1}$.
There is no substantial difference between this vibrational motion of the benzene molecule in liquid water or ammonia
and no higher-frequency vibrations of the solute than the CH stretches are present.
The bottom panel of Figure~\ref{fig:vibrations} shows the component of the IR absorption spectrum due to the solute--solute dipole moment correlations, with the aim of isolating and inspecting the solute-only contribution to the total IR spectrum.
The classical solute--solute self-terms (light blue and light red) contain the expected CH vibration peak at approximately 3200~cm$^{-1}$ that is consistent with the VDOS peak as well as the harmonic frequencies.
However, the spectra now contain new additional features at higher frequencies, too.
In water, these form a broad region of vibrational intensity between 3250 -- 4000~cm$^{-1}$ in the classical spectrum that can be interpreted as a double-peak structure with maxima located at 3700~cm$^{-1}$ and 3850~cm$^{-1}$.
With a quantum description, we can once again see NQEs in the form of a broadening and redshift of the spectral features to the point where the CH peak nearly fuses with the new features, almost leading to a single broad structure, where the original peaks are basically unrecognizable.
In contrast, in ammonia, the additional feature presents as a single sharp peak at 3625~cm$^{-1}$ under the classical approximation; the TRPMD IR spectrum of the ammonia solution was not calculated for reasons explained below.
Since these new features do not correspond to any peak in the solute VDOS, they cannot be ascribed to the vibrational motion of the atoms of the solute.
Rather, they must arise as an effect due to the solvent, which, through its own vibration, affects the dipole moment of the solute, leading to this feature in the solute--solute self-term spectrum.
This is consistent with the interpretation of the Raman experiment, which reports an equivalent feature in the SC spectrum and assigns it to the water molecule that forms a $\pi$-hydrogen bond.

\begin{figure}[tb!]
    \centering
    \includegraphics{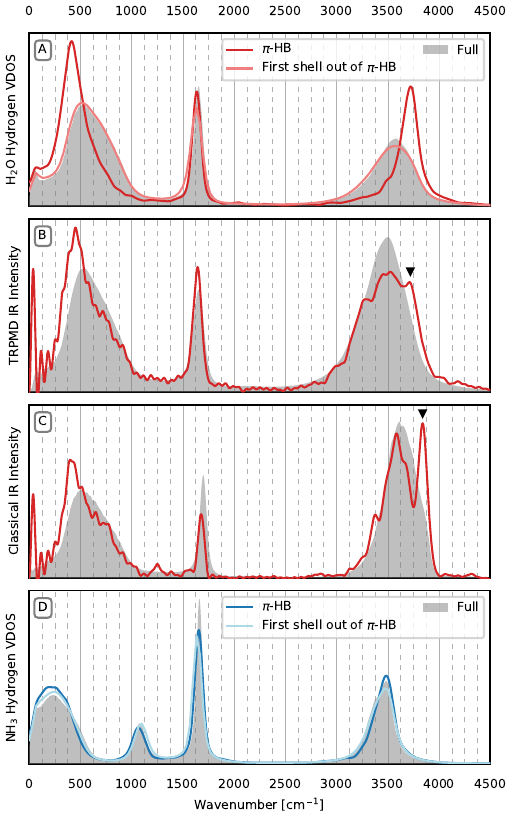}
    \caption{
    Vibrational characteristics of solvent molecules engaging in a $\pi$-hydrogen bond.
    Panel A: TRPMD VDOS of water hydrogen atoms for the full system (gray shading) and for selected hydrogens that engage in a $\pi$-hydrogen bond.
    In addition, the same quantity is shown for hydrogen atoms in the first solvent shell that do not engage in a $\pi$-hydrogen bond
    Panel B: TRPMD IR water self-term, again shown for the full system (gray shading) and the corresponding $\pi$-hydrogen bond subselection (red).
    The black triangle marks the relevant blueshifted feature.
    Note that the oscillation at the low-frequency end of the spectrum is a numerical artifact related to the practical execution of the Fourier transforms for the filtered spectra (see Section~\ref{si-sec:comp-details} of the Supporting information for details).
    Panel C: same data as in panel B, but obtained using classical mechanics.
    Note the pronounced split-peak feature in the OH stretch peak for molecules in a $\pi$-hydrogen bond.
    Panel D: same data as in Panel A, but for the liquid ammonia simulations.
    }
    \label{fig:resolved-vdos}
\end{figure}

However, the results presented until this point do not clarify whether the solvent-induced features in the solute-only IR spectrum are caused directly by solvent molecules that are in a $\pi$-hydrogen bond with the solute, as implied by the interpretation of the experiment.
To gain insight into the origin of the effect, we time-resolve the correlation functions that contribute to the \textit{solvent} VDOS and the whole IR spectrum and correlate them with the aforementioned $\pi$-hydrogen bond existence criteria.
This allows us to isolate the vibrational signature of molecules that are in a $\pi$-hydrogen bond at a given moment: this is possible given the lifetime of a $\pi$-hydrogen bond is much larger than the vibrational period.
The key results of this analysis for the simulations with quantum nuclei are summarized in Figure~\ref{fig:resolved-vdos}.

Let us first turn our attention to the aqueous solution.
The overall (unresolved) solvent hydrogen TRPMD VDOS is shown in panel A of Figure~\ref{fig:resolved-vdos} as the gray-shaded reference; it has the typical three-peak structure routinely observed for liquid water.~\cite{Morawietz2018/10.1021/ACS.JPCLETT.8B00133}
Restricting the presented VDOS to those solvent hydrogen atoms actively participating in a $\pi$-hydrogen bond shows a contribution that differs strongly from the total spectrum (Figure~\ref{fig:resolved-vdos}, panel A, dark red).
Notably, the maximum of the stretch band is blueshifted from 3575~cm$^{-1}$ to 3750~cm$^{-1}$, suggesting that the vibration in a $\pi$-hydrogen bond is stiffer than those in the bulk water--water hydrogen bond network.
In turn, this implies that the $\pi$-hydrogen bond vibration is more like a dangling or isolated OH bond, the vibration of which is also not softened by the presence of a relatively strong hydrogen bonding partner.
In contrast, the terahertz peak is shifted to lower frequencies, which is again consistent with the $\pi$-hydrogen bond being weaker than a water--water hydrogen bond and thus hindering rotation of the water molecule less.
To verify that the described VDOS shifts are specific for $\pi$-hydrogen bonds rather than a general effect of molecules in proximity to the solute, we also inspect the VDOS of solvent molecules that belong to the first solvent shell in terms of distance but are excluded by the $\pi$-hydrogen bond existence criteria.
We find that such molecules possess a vibrational signature essentially identical to the overall VDOS with no observable shifts (Figure~\ref{fig:resolved-vdos}, panel A, light red).

To explain these solvent-induced features in the solute-only IR component, we now turn our attention to the solvent--solvent self-term IR spectra.
The overall (unresolved) TRPMD water--water self-term IR spectrum is shown in panel B of Figure~\ref{fig:resolved-vdos} as a starting point, again in gray.
In this case, the $\pi$-hydrogen bond resolved counterpart brings around a more complicated structure of the OH stretch band (Figure~\ref{fig:resolved-vdos}, panel B, dark red) because, unlike VDOS, IR spectra cannot be resolved all the way to individual atoms and the effect of both OH-bond stretches is imprinted in the observed spectral feature.
As such, the faster vibration of the OH bond that participates in a $\pi$-hydrogen bond manifests as a high-frequency shoulder just below 3750~cm$^{-1}$ on the OH stretch peak (marked by a black arrow), while the rest of the dipole moment, including the second OH bond that points away from the solute and participates in the regular bulk hydrogen bond network, leads to a maximum at 3500~cm$^{-1}$.
The same split of the stretch peak due to different chemical environments of the two OH bonds is better visible in the classical IR spectrum (Figure~\ref{fig:resolved-vdos}, panel C, dark red), where a pronounced double-peak feature arises with maxima located at 3600~cm$^{-1}$ and 3850~cm$^{-1}$.
The frequencies of the observed double-peaks in both the classical and TRPMD case of the $\pi$-hydrogen bond resolved solvent-only IR spectrum correspond to those observed in the $\pi$-hydrogen bond resolved solute-only IR spectrum shown in Figure~\ref{fig:vibrations}.
This suggests that the effect observed in the solute-only IR spectrum is caused by the vibrating dipole of the solvent molecule inducing a dipole on the benzene solute at the frequency of this vibration.
Thus, the observed phenomenon in the solute-only spectrum is not specifically due to the $\pi$-hydrogen bond, but rather represents a general effect due to solute--solvent correlations.
However, since the interactions responsible for these correlations decay with distance and the $\pi$-hydrogen bonded molecule is the closest a solvent molecule can get to the solute (as demonstrated in Figure~\ref{fig:sdfs}), it is expected that the $\pi$-hydrogen bonded molecules have the strongest contribution to the effect; this interpretation is further supported by the fact that we found the imprint of the vibration of the distal OH bond, which does not directly participate in the formation of the $\pi$-hydrogen bond, in the solute-only IR spectrum as well.
The extent of participation of $\pi$-hydrogen bonded molecules in forming the solute-only spectral features could be directly tested, in principle, by calculating the solute-only spectrum in the absence of $\pi$-hydrogen bonds.
However, this is not technically possible with the present direct MD simulations: given the vibrational period, the characteristic length of the window where no $\pi$-hydrogen bonds exist is too short to be able to get a resolved spectrum corresponding to this state.
Simulations with the addition of a suitable biasing potential would be needed to obtain continuous trajectories free of $\pi$-hydrogen bonds.

The ammonia TRPMD VDOS (Figure~\ref{fig:resolved-vdos}, panel D) has one extra peak compared to the water spectra at around 1100~cm$^{-1}$ due to the umbrella vibration.
Interestingly, no frequency shifts are observed in the ammonia solutions after resolving the total solvent spectrum into the $\pi$-hydrogen bond and non-$\pi$-hydrogen bond first-shell contributions: all spectra are essentially identical.
This result has a simple physical explanation: unlike in water, where the bulk hydrogen bond structure is strong, in ammonia, it is nearly non-existent, and, as such, the molecule in a $\pi$-hydrogen bond does not differ substantially from the other solvent molecules in the bulk.
For this reason, the ammonia counterpart of the solvent-induced peak in the solute-only spectrum (Figure~\ref{fig:vibrations}, bottom panel in blue) is represented by a single narrow peak and not a complicated double-peak structure as it is in water.

In summary, the vibrational character of molecules engaging in $\pi$-hydrogen bonds is different in both solvents, but this difference is due to the difference in the hydrogen bond strength within the respective solvent bulks, not a difference in the nature of the $\pi$-hydrogen bonds themselves.

\section{Conclusions}

The reported TRPMD simulations of solutions of benzene in liquid water and ammonia provide a detailed computational perspective on the phenomenon of $\pi$-hydrogen bonding.
To this end, we relied on the combination of accurate \textit{ab initio} electronic structure with high computational efficiency in the form of C-NNPs to converge the statistical properties of interest.
In this work, we found clear evidence of $\pi$-hydrogen bonding in the solvation structure of benzene in both solvents and characterized it in several complementary ways.
We discussed the orientational flexibility of $\pi$-hydrogen bonded solvent molecules, described the strength of the interaction by its equilibrium population, showed evidence of its anticooperativity in water, and determined its lifetimes in both solvents.
Furthermore, we modeled the VDOS and IR spectra of these systems and, employing a spatial and temporal decomposition that is only available in a computational approach, found and explained imprints of the $\pi$-hydrogen bonding interaction that are consistent with previous experimental findings in water.
Our simulations point to the fact that $\pi$-hydrogen bonding is a prominent interaction present in the studied systems most of the time.
This suggests relevance for situations where aromatic species come into contact with water, such as in the solvation of biological residues; some authors even suggest a possible role of $\pi$-hydrogen bonding in biological scenarios such as signaling~\cite{Gierszal2011/10.1021/JZ201373E}.
An accurate description by empirical force fields would thus be instrumental, yet it is currently not available~\cite{Allesch2008/10.1063/1.2806288/928000}.
Our work provides insight as well as an extensive \textit{ab initio} reference for the future development of adequate force fields.

The question of why $\pi$-hydrogen bond anticooperativity presents a challenge for machine learning potentials, as identified by our \textit{ab initio} simulations, remains open.
Rectifying this issue will provide insight into the functionality and shortcomings of these potentials and will enable accessing other hard-to-converge aspects of $\pi$-hydrogen bonding, such as the further details of the thermodynamics of the formation of this bond.
An additional level of improvement will be achieved by employing machine-learned models for predicting molecular dipoles and polarizabilities~\cite{Lewis2023/10.1063/5.0154710/2900715}, which will give access to better converged IR spectra and open the doors to computationally accessible Raman spectroscopy as well.

\begin{acknowledgments}

O.M. thanks Dor Ben-Amotz for stimulating discussions of vibrational spectroscopy of $\pi$-hydrogen bonds in solution.
The authors acknowledge support from the Czech Science Foundation, project No. 21-27987S.
This work was supported by the Ministry of Education, Youth and Sports of the Czech Republic through the e-INFRA CZ (ID:90254).

\end{acknowledgments}

\end{bibunit}


\clearpage

\setcounter{section}{0}
\setcounter{equation}{0}
\setcounter{figure}{0}
\setcounter{table}{0}
\setcounter{page}{1}

\renewcommand{\thesection}{S\arabic{section}}
\renewcommand{\theequation}{S\arabic{equation}}
\renewcommand{\thefigure}{S\arabic{figure}}
\renewcommand{\thepage}{S\arabic{page}}
\renewcommand{\citenumfont}[1]{S#1}
\renewcommand{\bibnumfmt}[1]{$^{\rm{S#1}}$}

\title{Supporting information for: \mytitle}
{\maketitle}

\begin{bibunit}

\section{Computational details}
\label{si-sec:comp-details}

\subsection{AIMD simulations}

An initial configuration of the benzene--water system was obtained by placing the benzene solute into a sufficiently large simulation box and randomly solvating it with 64 water molecules.
Due to the unavailability of experimental densities of benzene--water solutions, such a structure was pre-equilibrated using an empirical force field for 1 ns under the isothermal--isobaric ensemble at the temperature of 223~K and the pressure of 1~atm to equilibrate the cell volume using the Gromacs program~\cite{Pall2020/10.1063/5.0018516/199476}.
The equilibrium side length of the box was estimated to be 12.667~\AA.
The resulting structure was additionally pre-equilibrated for 3~ps using the generalized-gradient-approximation revPBE-D3~\cite{Perdew1996/10.1103/PhysRevLett.77.3865, Zhang1998/10.1103/PhysRevLett.80.890, Goerigk2011/10.1039/c0cp02984j} density functional with the TZV2P basis set under canonical conditions (SVR thermostat~\cite{Bussi2007/10.1063/1.2408420/186581} with a 50~fs time constant) at 300~K using the CP2K software~\cite{Kuehne2020/10.1063/5.0007045/199081} (version 7.1) to drive the integration of the dynamics with an integration time step of 0.5~fs.
We relied on the Quickstep module~\cite{VandeVondele2005} of CP2K for efficient DFT electronic structure calculations based on the Gaussian and Plane Waves (GPW) approach using a plane-wave cutoff of 600~Ry.
The sequence continued by changing the density functional to the final production-level hybrid revPBE0-D3~\cite{Perdew1996/10.1103/PhysRevLett.77.3865, Zhang1998/10.1103/PhysRevLett.80.890, Adamo1999/10.1063/1.478522, Goerigk2011/10.1039/c0cp02984j} and equilibrating for an additional 3~ps under the same conditions, using a local SVR thermostat with a 50~fs time constant to enforce an efficient thermalization to the target temperature of 300~K.
From then on, we switched the local thermostat used for equilibration to a global SVR thermostat~\cite{Bussi2007/10.1063/1.2408420/186581}, extended its time constant to 200~fs, and continued with production MD for the total simulation time of 20~ps.
We have repeated these steps for 5 uncorrelated initial conditions, yielding 100~ps of statistical sampling in 5 20~ps-long AIMD trajectories.
During the parts of the calculations involving the revPBE0-D3 functional, we have relied on the auxiliary density matrix method with the cpFIT3 fitting basis set~\cite{Guidon2010/10.1021/ct1002225} as well as on the Schwartz inequality to screen small enough elements of the density matrix~\cite{Guidon2008/10.1063/1.2931945} in combination with the always-stable predictor-corrector method~\cite{Kolafa2005} to extrapolate the Kohn--Sham wave function to accelerate the computationally challenging simulation.
An analogous approach was taken for the classical benzene--ammonia simulations, in this case fixing the box length to 13.855~\AA\ at the simulated temperature of 223~K.
Exhaustive details of the benzene--ammonia classical simulations are discussed in our previous work~\cite{Brezina2020/10.1021/ACS.JPCLETT.0C01505}, which introduced these simulations for the first time.

The path-integral simulations were realized by combining the CP2K software to evaluate the DFT electronic energies and forces with the i-PI program~\cite{Kapil2019/10.1016/J.CPC.2018.09.020} to drive the TRPMD simulations~\cite{Rossi2014/10.1063/1.4883861/73314}, relying on communication through a socket interface.
An identical simulation protocol was employed for both water- and ammonia-containing systems.
In this case, we set out from the above-mentioned classical 3~ps pre-equilibrated structures at the hybrid DFT level and continued with another 1~ps of equilibration while employing a 64-replica path-integral representation of the atomic nuclei.
To lower the computational requirements of the \textit{ab initio} PIMD simulations, we have relied on the ab initio ring polymer contraction (RPC) approach~\cite{Markland2008/10.1063/1.2953308/842694, Marsalek2016/10.1063/1.4941093} and employed a multiple-time step integration scheme~\cite{Tuckerman1992/10.1063/1.463137}.
This allowed us to use the full, hybrid revPBE0-D3 calculation on the centroid mode with a longer 2.0~fs time step while treating the intermittent motion of the individual replicas at the computationally much more manageable semi-empirical self-consistent charge density-functional tight-binding~\cite{Porezag1995/10.1103/PhysRevB.51.12947} (SCC-DFTB) level with a 0.25~fs time step.
For equilibration purposes, we used the local path-integral Langevin thermostat~\cite{Ceriotti2010/10.1063/1.3489925} (PILE-L) with a 100~fs time constant to ensure the desired temperature of 300~K.
After this initial equilibration period of the path-integral replicas, we proceeded to the production simulation, in which we changed the PILE-L thermostat to the global formulation of the path-integral Langevin thermostat~\cite{Ceriotti2010/10.1063/1.3489925} (PILE-G), with a time constant of 100~ps and damping scaling parameter $\lambda$ set to 0.5 to ensure TRPMD conditions for the propagation.
Such simulation was continued until the total simulation time of 50~ps was reached, collecting both replica and centroid positions and velocities each 2.0~fs.
In total, we thus obtained 250~ps of path-integral sampling in five continuous trajectories.

\subsection{Simulation of IR spectra}

As clarified in Equation~\ref{eq:ir} of the main text, the simulation of the IR spectrum requires the knowledge of instantaneous molecular dipoles along the trajectory.
In the classical case, this was achieved using an approach based on Wannier localization of molecular orbitals performed on the fly, directly yielding molecular dipoles as a simulation output.
In the path-integral case, this is not an option due to using RPC: individual replicas are never treated at the full DFT level.
Therefore, the molecular dipoles were obtained through the post-processing of the centroid trajectories to obtain the time-dependent molecular dipoles.
This was done despite the fact that the dipole operator is not exactly linear in atomic positions, and therefore, one cannot, strictly speaking, use the centroid-based time-dependent dipoles as time-dependent observables for the RPMD approximation to time correlation functions~\cite{Craig2005/10.1063/1.1777575}.
However, we show below in Section~\ref{si-sec:additional_results} that this approximation is very reasonable for practical purposes while offering substantial computational savings.

To obtain both VDOS and IR spectra presented and discussed in the main text, we have relied on in-house code implemented in the Python programming language to calculate the relevant discretized time correlation functions from the simulated time series and to perform the fast Fourier transforms (FFTs) into the frequency domain.
Regarding the numerical aspects of the FFTs, we applied a 3~ps-wide Hann apodization window to control the noise in the resulting spectra and padded the signal with a 10~ps long zero-valued tail to increase the numerical resolution of the resulting curves.

\subsection{Machine Learning Potentials}

\begin{figure}[tb!]
    \centering
    \includegraphics[width=\linewidth]{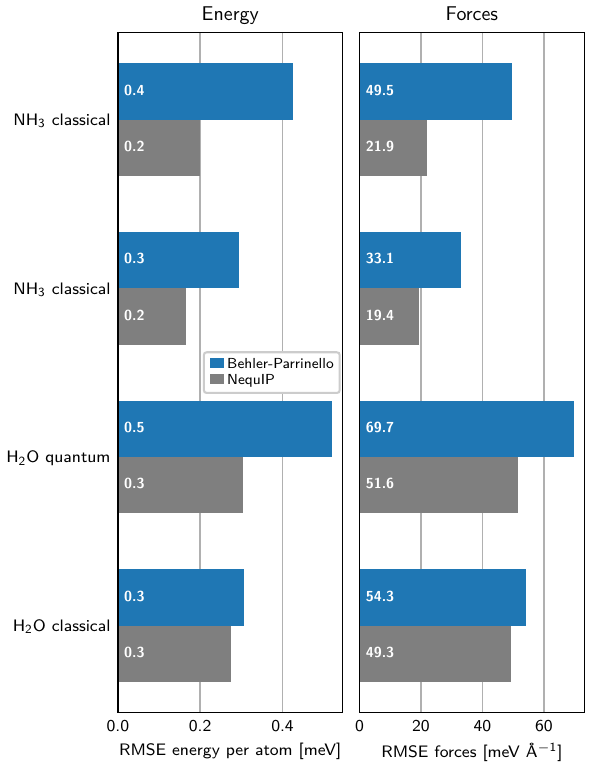}
    \caption{
        Energy RMSE per atom and Force RMSE for the Behler-Parrinello C-NNPs and NequIP NNPs for benzene solvated in water or ammonia.
    }
    \label{si-fig:rmse}
\end{figure}

We obtained four training data sets calculated at the revPBE0-D3 level of theory using the query by committee (QbC) method~\cite{Seung1992/10.1145/130385.130417, Artrith2012/10.1103/PhysRevB.85.045439} relying on the workflow outlined in our previous work~\cite{Schran2020/10.1063/5.0016004}.
The candidate structures for the QbC process stem from the original \textit{ab initio} revPBE0-D3 AIMD trajectories of benzene solvated in water or ammonia, simulated with either classical or quantum treatment of the nuclei.
From these large sets of candidates, we added the 10 structures with the highest force disagreement in every iteration of QbC to construct data sets of 400 structures each.
8 individual Behler--Parrinello NNPs~\cite{Behler2007/10.1103/PHYSREVLETT.98.146401} were trained on each of the training data sets for 500 epochs using the N2P2 code to form the final committee of NNPs (C-NNP) used for MD simulations.
Furthermore, equivariant message-passing graph NNPs using the NequIP architecture~\cite{Batzner2022/10.1038/s41467-022-29939-5} were also trained on the same training data sets.
While the training geometries were selected during QbC to work well with Behler--Parrinello NNPs, the data efficiency of NequIP ensures dependable predictions even with the non-optimized training data.
Both MLP architectures were tested on independent test sets for the 4 different systems containing 200 structures each, which were randomly selected from the \textit{ab initio} trajectories.
The results from these tests, namely the energy root mean square error (RMSE) per atom and the RMSE of the force components, can be seen in Figure~\ref{si-fig:rmse}.
It shows that all 8 MLPs perform well for their respective systems and that the NequIP NNP consistently outperforms the Behler--Parrinello C-NNPs, especially for the ammonia systems and the structures originating from path integral MD, despite the fact that the training sets were not optimized for NequIP.
In addition, we performed AML scoring calculations~\cite{Schran2021/10.1073/PNAS.2110077118}, finding the scores of $> 98$\% for radial distribution functions and $\sim 90$\% for VDOS, consistent with the quality of previously employed models.

\subsection{Simulations with MLPs}

With the optimized MLPs at hand, we ran MD simulations of both systems to improve sampling statistics compared to the \textit{ab initio} case.
In the case of the classical description of nuclei, the only component that changed was the employed potential from the explicit hybrid DFT to the corresponding MLP; other simulation settings were kept to achieve maximum consistency.
With this setup, we obtained 4 trajectories, each 500~ps-long, giving us a total of 2~ns of simulated time.
Along the trajectories, we have collected atomic positions every 10~fs and atomic velocities every 2~fs to make it possible to calculate VDOS.
In the TRPMD simulations, the computational efficiency of the MLPs allowed us to drop the RPC and to treat the fully resolved, 64-replica ring polymers at the level of the MLP.
Therefore, these simulations can be considered of higher quality than the original \textit{ab initio} ones as they contain one level of approximation less.
In turn, the consistency between the MLP- and DFT-based TRPMD simulations validates the initially employed RPC, as also done previously in our NNP-based path-integral simulations of liquid water~\cite{Schran2020/10.1063/5.0016004}.
Consistently with the classical case, we have obtained 2~ns of total simulated time in 4 continuous TRPMD trajectories.
In this case, we collected replica positions only every 300~ps (no replica velocities were collected), centroid positions every 10~ps and centroid velocities every 2~ps.

\section{Additional results}
\label{si-sec:additional_results}

In the paragraphs below, we present additional results that support the discussion in the main text.

\subsection{Normalized water $\pi$-hydrogen bond orientation distribution}

\begin{figure}[tb!]
    \centering
    \includegraphics[width=\linewidth]{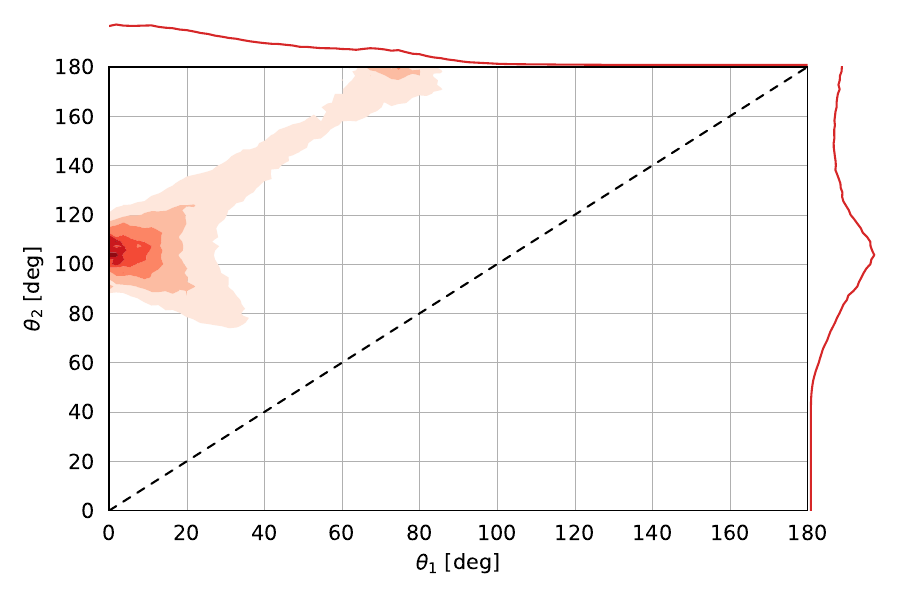}
    \caption{
    Plot of probability density $p'(\theta_1, \theta_2)$ for the benzene--water system.
    The colormap bears the same meaning as in Figure~\ref{fig:orientation} in the main text: the darker the color, the higher the probability density value.
    The side panels show the marginal distributions in $\theta_1$ and $\theta_2$.
    }
    \label{si-fig:sine}
\end{figure}

Figure~\ref{fig:orientation} of the main text shows distributions of the angles $\theta_j$ between individual XH bonds of the solvent molecules inside the selection cylinder and the $z$-axis.
As with any other probability distribution in angular coordinates, this is affected by the orientational entropy that favors states with angles closer to $90^\circ$ due to the number of possibilities of realizing those states in comparison to the single possibility of realizing the state with $0^\circ$: the number of realizations is proportional to the sine of the angle.
Figure~\ref{si-fig:sine} shows the distribution 
\begin{equation}
    p'(\theta_1, \theta_2)
    =
    \frac{p(\theta_1, \theta_2)}{\sin{(\theta_1)} \sin{(\theta_2)}},
\end{equation}
where $p(\{\theta_j\})$ is the original probability density shown in Figure~\ref{fig:orientation}.
This clearly shows that once the effect of the orientational freedom is lifted, the $0^\circ$ configuration featuring an ideally vertical $\pi$-hydrogen bond becomes favored.

\subsection{Cooperativity from C-NNP trajectories}

\begin{figure}[tb!]
    \centering
    \includegraphics[width=\linewidth]{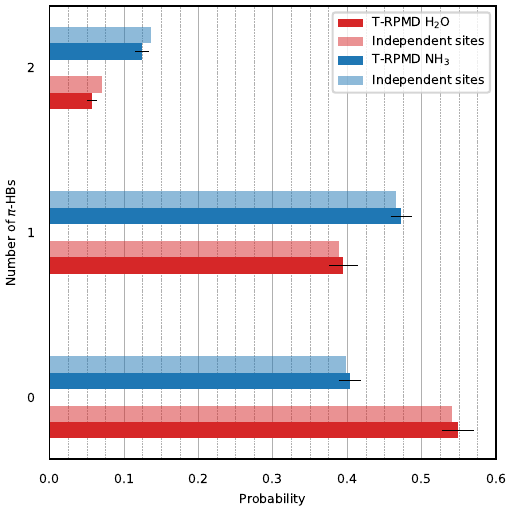}
    \caption{
    Probabilities of finding the zero-, one- and two-bonded configurations within our simulated trajectories using the Behler--Parrinello C-NNP.
    The darker bars show real, observed populations.
    The lighter bars then show probabilities that are calculated based on the premise of binding site independence.
    }
    \label{si-fig:cnnp-populations}
\end{figure}

In Figure~\ref{si-fig:cnnp-populations}, we present the analogous plot to the one in Figure~\ref{fig:populations} of the main text, but obtained with the C-NNP MLPs.

\subsection{Changes in electron density in benzene--solvent gas phase clusters}

\begin{figure}[tb!]
    \centering
    \includegraphics[width=\linewidth]{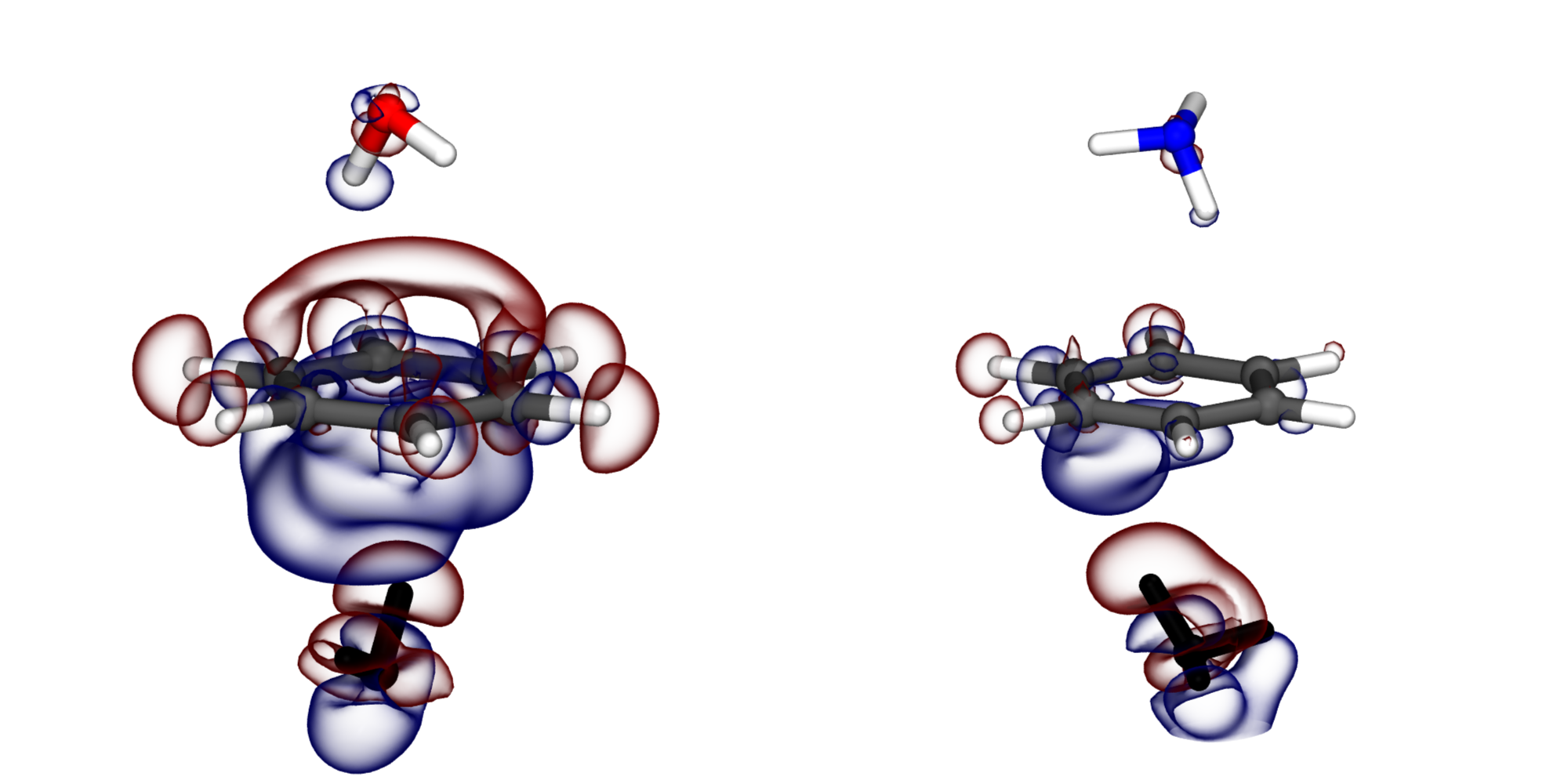}
    \caption{
    Spatial plots of $\Delta \rho(\mathbf{r})$ (see text for definition) in benzene--ammonia (left) and benzene--water (right) shown at the $\pm 3.5 \cross 10^{-4}$~$a_0^{-3}$ contour (red for $+$, blue for $-$).
    }
    \label{si-fig:delta-rho}
\end{figure}

To gain qualitative insight into the observed anticooperativity of $\pi$-hydrogen bonds at the gas-phase level, we optimized microsolvated clusters consisting of a single benzene and two $\pi$-hydrogen bonded solvent molecules (Figure~\ref{si-fig:delta-rho}) at the revPBE0-D3BJ/def2-TZVP level of theory using the QChem software~\cite{Epifanovsky2021/10.1063/5.0055522/1074802}.
Then, using these geometries, we have calculated the following electron densities: $\rho_\mathrm{full}(\mathbf{r})$ for the full systems, and $\rho_\mathrm{A}(\mathbf{r})$ for the same geometry with one of the solvent molecules removed (shown in black in Figure~\ref{si-fig:delta-rho}) and, finally, $\rho_\mathrm{B}(\mathbf{r})$ of the removed solvent molecule at its original coordinates with the rest of the system removed.
Then, we calculated a difference density
\begin{equation}
    \Delta \rho(\mathbf{r})
    =
    \left[\rho_\mathrm{A}(\mathbf{r}) + \rho_\mathrm{B}(\mathbf{r})\right] - \rho_\mathrm{full}(\mathbf{r}),
\end{equation}
which is the quantity shown in Figure~\ref{si-fig:delta-rho} as transparent contours.
Clearly, this has both positive and negative values (red and blue, respectively) that denote the enrichment or depletion of electron density at a given point in space after the vertical removal of the solvent molecule.
This polarization is substantially stronger in the case of water (Figure~\ref{si-fig:delta-rho}, left) than in the case of ammonia (Figure~\ref{si-fig:delta-rho}, right).
It is likely that the quantitative difference in the two solvents, in combination with the influence of the rest of the condensed phase, is responsible for the overall anticooperativity effect observed in water and the lack of it in ammonia.

\subsection{Validation of the centroid approximation for the benzene molecular dipole moment}

\begin{figure}[tb!]
    \centering
    \includegraphics[width=\linewidth]{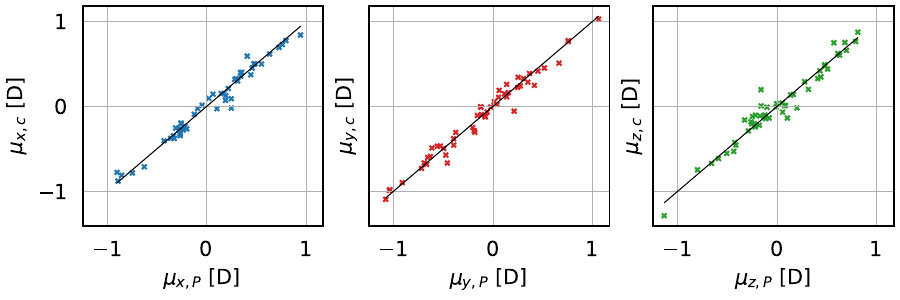}
    \caption{
    The correlation of the components of the benzene dipole averaged over the replicas compared to the dipole of the centroid. 
    }
    \label{si-fig:corr-dipole}
\end{figure}

Finally, we present arguments for using the centroid approximation in the calculations of IR spectra from TRPMD simulations.
Correctly, the TRPMD time-dependent dipole observable would be calculated as
\begin{equation}
    \pmb{\mu}_P[\mathbf{R}(t)]
    =
    \frac{1}{P} \sum_{k=1}^P \pmb{\mu}[\mathbf{R}_k(t)],
\end{equation}
which requires the knowledge of the dipole values evaluated for individual replicas with geometries $\mathbf{R}_k$.
This makes the correct calculation of IR spectra from path-integral calculations a potential computational challenge.
The dipole in general is not a linear function of nuclear positions due to its non-trivial dependence on electronic structure and, therefore,
\begin{equation}
    \pmb{\mu}_P[\mathbf{R}(t)]
    \neq
    \pmb{\mu}[\mathbf{R}_\mathrm{c}(t)],
\end{equation}
where
\begin{equation}
    \mathbf{R}_\mathrm{c}(t)
    =
    \frac{1}{P} \sum_{k=1}^P \mathbf{R}_k(t)
\end{equation}
is the ring-polymer centroid.
However, in practice the dipole often exhibits a quasi-linear behavior with respect to atomic positions and one can use the centroid-based estimate of the dipole observable as a good approximation,
\begin{equation}
    \pmb{\mu}[\mathbf{R}_\mathrm{c}(t)]
    \approx
    \pmb{\mu}_P[\mathbf{R}(t)].
\end{equation}
This is much simpler from a computational perspective because one only has to evaluate dipoles on the classical-like, centroid trajectory.
In Figure~\ref{si-fig:corr-dipole}, we demonstrate on the example of the molecular dipole of benzene in the benzene--water TRPMD simulations that the centroid- and replica-based dipoles exhibit a reasonable correlation.
For this reason, we have used such a centroid approximation in our calculation of path-integral IR spectra.

\end{bibunit}

\end{document}